\documentclass[12pt,epsf,amstex]{article}
\usepackage [dvips]{graphicx}
\usepackage{amsmath}
\usepackage{amssymb}
\usepackage{psfig}
\usepackage{epsfig}

\addtocounter{secnumdepth}{1}
\newcommand{\vecR}{{\vec{R}}}
\newcommand{\vecS}{{\vec{S}}}

\newcommand{\half}{\frac{1}{2}}
\newcommand{\summ}{\sum_{m=1}^n}

\newcommand{\dd}{\mbox{d}}
\newcommand{\ee}{\mbox{e}}

\newcommand{\la}{\langle}
\newcommand{\ra}{\rangle}
\newcommand{\rao}{\rangle_{_0}}

\newcommand{\beq}{\begin{equation}}
\newcommand{\eeq}{\end{equation}}
\newcommand{\bea}{\begin{eqnarray}}
\newcommand{\eea}{\end{eqnarray}}
\def\lsim{\:\raisebox{-0.5ex}{$\stackrel{\textstyle<}{\sim}$}\:}
\def\gsim{\:\raisebox{-0.5ex}{$\stackrel{\textstyle>}{\sim}$}\:}

\begin{document}

\thispagestyle{empty}

\title{\sf The perimeter of large planar Voronoi cells: \\[1.5mm] 
a double-stranded random walk}

\author{{\bf H.\,J. Hilhorst}\\[5mm]
{\small Laboratoire de Physique Th\'eorique}\\[-1mm]
{\small B\^atiment 210, Universit\'e de Paris-Sud}\\[-1mm]
{\small 91405 Orsay Cedex, France}\\}

\maketitle

\begin{abstract}
Let $p_n$ be  the probability 
for a planar Poisson-Voronoi cell to have exactly $n$ sides.
We construct the asymptotic expansion of $\log p_n$
up to terms that vanish as $n\to\infty$. Along with it comes a nearly
complete understanding of the structure of the large cell.
We show that {\it two independent biased random walks} executed by
the polar angle
determine the trajectory of the cell perimeter.
We find the limit distributions of (i) the angle between two successive
vertex vectors, and (ii) the one between two successive perimeter segments.
We obtain the probability law for the perimeter's long wavelength
deviations from circularity. 
We prove Lewis' law and show that it has coefficient $1/4$.{}
\end{abstract}


\noindent {\bf 1. Introduction}
\vspace{2mm}

Cellular structures are observed in a great diversity of situations 
in science and technology.
A common phenomenological description of such
structures makes use of Voronoi cells \cite{Voronoi08}, 
constructed around centers which for simplicity
are referred to as ``particles'': the Voronoi cell of a
particle $a$ is made up of those
points in space for which $a$ is the closest particle.
Hence Voronoi cells constitute a ``tessellation'', 
{\it i.e.}~a division of space into nonoverlapping regions.
An excellent review of tessellations,
with in its introduction many historical and anecdotal 
details, is due to Okabe {\it et al.} \cite{Okabeetal00}.
It offers a thorough discussion of much
fundamental and applied work in this area.
This includes applications to biological tissues, mesh generation 
in numerical algorithms, coding in telecommunication, 
{\it etc.} 

Whereas the
Wigner-Seitz cell associated with the regular lattices
of solid state physics is a special case, the generic
Voronoi tessellation is based on randomly distributed particles;
for uniformly distributed point-particles 
the result is called a {\it Poisson-Voronoi\,} tessellation. 

Voronoi cells appear in physics
for many different reasons.
Meijering \cite{Meijering53} used them to characterize the end result of 
a crystal growth process starting from independent centers.
In the theory of the solid-liquid transition they serve to
define nearest-neighbor relations between atoms, which in turn allow the
identification of lattice defects. 
Recently Voronoi tessellations of fractal sets have been 
considered \cite{HS98}. 
In field theory  the use of random lattices and the associated 
Voronoi tessellation 
\cite{Christetal82,ID89} was motivated by their
advantage of being statistically invariant under
arbitrary translations and rotations while still preserving
a short-distance cutoff.

Here we are interested in two-dimensional 
Poisson-Voronoi tessellations.
Planar Voronoi cells are convex polygons.
The first and foremost quantity that appears in both theoretical and
applied work is the probability $p_n$ for a planar Voronoi cell to be
$n$-sided. Approximate values for $p_n$ have been tabulated by many
authors \cite{Okabeetal00}; 
they peak at $n=6$ and fall off rapidly for larger $n$.
Connected questions 
concern the statistics of the cell 
area, its perimeter, and its angles, as well as
correlations with neighboring cells.
They have been studied by methods ranging from
heuristic arguments and Monte Carlo
simulations to exact methods. 
Whereas some statistical
properties are readily obtained,
others, among which the fraction $p_n$, 
still defy solution \cite{ID89,Okabeetal00}.

Interest in planar Voronoi cells having a {\it large number} 
$n$ of sides first arose in biology.
Lewis' empirical law \cite{Lewis,Okabeetal00},
formulated more than seventy years ago,  
states that when $n$ becomes large, the average area of the
$n$-sided cell grows as $\simeq c(n-2)/\nu$, where $\nu$ is the particle
density.  
In spite of attempts, this asymptotic
proportionality with $n$ has not so far been proved \cite{Rivieretc}.
Numerical estimates of the coefficient $c$ are in the range $0.199-0.257$
\cite{DI84,Lecaeretc,Okabeetal00}.
In metallography large cells were observed by Aboav \cite{Aboav70} in
the arrangement of grains in a polycrystal. His
empirical law states that the neighbor of an $n$-sided cell
has itself on average \,$a_0+a_1 n^{-1}$\, neighbors. 
Only recently Hug {\it et al.} \cite{Hugetal04} showed rigorously
that when a Voronoi cell becomes large (in an appropriate sense), 
its shape tends towards a circle.
In a different approach mathematicians study
those large cells that allow for an inscribed circle 
whose given radius becomes asymptotically large \cite{Calka02a}.

Drouffe and Itzykson (DI) \cite{DI84}
estimated the probability ${p}_n$ for a planar Voronoi cell to be $n$-sided
by Monte Carlo simulations for $n$ up to $50$.
The rejection rate limits the statistical precision of generating such
very rare events.
All other simulations cited in Ref.\,\cite{Okabeetal00}
(see  {\it e.g.} \cite{Lecaeretc}) were restricted to
$n\lsim 14$, so that the work by DI is still today the main
reference for the large $n$ behavior. 
\vspace{2mm}

This note is the first announcement of a collection of new analytical
results on planar Poisson-Voronoi tessellations.
All of them are by-products of
our asymptotic evaluation of $p_n$, and
together they lead to a nearly complete understanding of the large 
$n$-sided cell.
Our analysis uses an accumulation of classical mathematical
methods; however, the asymptotics is beset by a great number of
difficulties, of which the correct choice of variables of
integration is merely the first one. 
It will be possible here to indicate only the main steps
and the main results.

We begin by a precise statement of the question.
Let a subdomain of the plane, of area $L^2$, contain
${N}$ particles with tags $a$,
placed at positions $\vec{\mathfrak{R}}_a$
chosen independently and with  uniform probability,
and let ${N},{L}\to\infty$  at fixed ${N/L^2}=\nu$.
We consider the Voronoi cell 
of an arbitrarily chosen particle $a_0$ whose position will be
the origin of the coordinate system.
The perpendicular
bisectors of the $\vec{\mathfrak{R}}_a$ (with $a\neq a_0$)
pass through the 
{\it midpoints} $\vecR_a=\frac{1}{2}\vec{\mathfrak{R}}_a$,
which are uniformly distributed with density $4\nu$. 
Each side of the cell
is a segment of one of the perpendicular bisectors.
An $n$-sided cell may be fully specified by $n$
mid\-point vectors 
$\vecR_m$ ($m=1,\ldots,n$);
or alternatively by the $n$
{\it vertex vectors} $\vecS_m$ ($m=1,\ldots,n$) that connect 
the origin to its vertices.

The probability $p_n$ has been
expressed \cite{DI84,ID89,Calka02c} as the $2n$-dimensional integral
\begin{equation}
{p}_n=\frac{1}{n!}
\int \Big[ \prod_{m=1}^n \dd \vecR_m \Big]
\chi\,\, \ee^{-{\cal A}}\,,
\label{exprR2pn}
\end{equation}
where we scaled lengths such that $4\nu=1$; it is analogous to a
partition function in statistical mechanics.
The functions $\chi$ and ${\cal A}$ enforce two
conditions necessary for the $n$ 
vectors $\vecR_m$ to define a valid Voronoi cell. 
First, each of the $n$ associated 
bisectors must contribute a nonzero segment to
the perimeter. Calka \cite{Calka02c}
formulates this condition as a set of $n$
inequalities, and 
$\chi$ is the indicator function of the phase space domain
where they are satisfied.
Secondly, the union of the $n$ disks centered at 
$\frac{1}{2}\vecS_m$ and of radius $\frac{1}{2}S_m$ 
must be void of midpoints. 
The probability for this is $\ee^{-{\cal A}}$,
where ${\cal A}$ is the area of the void region.
Explicit expressions for ${\cal A}$ 
were given by DI \cite{DI84} 
and by Calka \cite{Calka02c}. Eq.\,(\ref{exprR2pn}) 
has been rewritten and
evaluated to seven decimal places \cite{HayenQuine00}
for $n=3$, but has otherwise remained untractable
analytically.
\vspace{5mm}

\noindent {\bf 2. The perimeter as a Markov process}
\vspace{2mm}

Our approach originated from an attempt to describe
the perimeter of a Voronoi cell
as a Markov process with the polar angle as
independent variable. 
We order the midpoints
$\vecR_m\equiv(R_m,\Phi_m)$ such that 
$0<\Phi_1<\ldots<\Phi_{n-1}<\Phi_n\equiv 2\pi$ and 
denote by
$\vecS_m\equiv(S_m,\Psi_m)$ the vertex where the bisectors through 
$\vecR_{m-1}$ and $\vecR_m$ intersect.
Let $\beta_m$ be the angle
between $\vecR_{m-1}$ and $\vecS_{m}$, 
and $\gamma_m$ the one between $\vecS_{m}$ and $\vecR_m$. 
These angles may be positive or negative, 
since a midpoint need not (and for large $n$ generically will not!)
itself be part of the side that it defines. 
The Markov process description leads to a transition probability
between two successive midpoints which has, essentially,
the form $R_m^2 T_m \ee^{-{\cal A}_m}$. Here
${\cal A}_m$,
which satisfies ${\cal A}=\summ{\cal A}_m$,
is the area of a region delimited by three circular arcs that should be
void of midpoints if the
perimeter is to travel uninterruptedly between $\vecS_{m-1}$ and
$\vecS_m$; 
and $T_m={\cos\gamma_m\sin(\beta_m+\gamma_m)}/{\cos^3\beta_m}$ is the
(unnormalized) probability density for the perimeter to turn
by $\beta_m+\gamma_m$ at vertex $\vecS_m$.
Considerable algebra then gives
\bea
{p}_n &=& \frac{1}{n}\int\! \dd\beta\dd\gamma\,\, 
\delta\big(\sum_{m=1}^n(\beta_m+\gamma_m)-2\pi\big)\,
\delta(G)\nonumber\\
&& \times\,
\Big[ \prod_{m=1}^n \rho_m^2 T_m \Big]\,\, 
\int_0^\infty \!\dd R\, R^{2n-1} \ee^{-\pi R^2 A(\beta,\gamma)}\,,
\label{exprbgpn}
\eea
which calls for several explanations. 
(i) We abbreviated
\bea
\int\dd\beta\dd\gamma &=&
\int_{-\pi/2}^{\pi/2}\!\!\dd\beta_1
\int_{-\beta_{_1}}^{\pi/2}\!\!\dd\gamma_1 
\int_{-\gamma_{_1}}^{\pi/2}\!\!\dd\beta_2
\int_{-\beta_{_2}}^{\pi/2}\!\!\dd\gamma_2 \ldots \nonumber\\
&& \phantom{xxxx}
\ldots\int_{-\gamma_{_{n-1}}}^{\pi/2}\!\!\dd\beta_n 
\int_{-\beta_{_n}}^{\pi/2}\!\!\dd\gamma_n\,\,     
\theta(\gamma_n+\beta_1)\,. 
\label{defintbetagamma}
\eea
Because of how they are nested, these integrations 
impose an orientation on the perimeter.
(ii) For an arbitrary set of angles 
$(\beta,\gamma)\equiv\{\beta_m,\gamma_m\}$ 
the perimeter will spiral instead of
close onto itself after a full turn of $2\pi$;
the factor $\delta(G)$, where
$G = (2\pi)^{-1} \sum_{m=1}^n \log(\cos\gamma_m/\cos\beta_m )$,
enforces its closure.
(iii) The ratios $\rho_m\equiv R_m/R$, where $R$ is the average midpoint
distance $R=n^{-1}\sum_{m=1}^n R_m$,
are functions of the angles defined by
\beq
\rho_m=\frac{\cos\gamma_m\cos\gamma_{m-1}\ldots\cos\gamma_1}
{\cos\beta_m\cos\beta_{m-1}\ldots\cos\beta_1}\, \rho_n
\label{condcosbcosg}
\eeq
for $m=1,\ldots,n-1$, together with
the sum rule $n^{-1}\sum_{m=1}^n\rho_m=1$.
(iv) Finally, we set ${\cal A}=\pi R^2A(\beta,\gamma)$.

Eq.\,(\ref{exprbgpn}) fully defines
our starting point.
As a check one may derive it    
directly from (\ref{exprR2pn}) by a coordinate 
transformation, in which the factors $\rho_m^2 T_m$ then
appear as the Jacobian. The decisive
advantage of (\ref{exprbgpn}) over (\ref{exprR2pn}) is that
the integration limits have been made explicit 
so that the function $\chi$ is no longer needed.

Integrating on $R$ in (\ref{exprbgpn}) yields
$\frac{1}{2}(n-1)!\,[\pi A(\beta,\gamma)]^{-n}$.
If, as will appear,
$A(\beta,\gamma)$ differs negligibly
from unity for $n\to\infty$, 
a steepest descent analysis of the same integral 
shows that it draws its main contribution from $R$ 
within a width of order $n^0$ around the saddle point 
$R_{\rm c}=\sqrt{n/\pi}$.

Let now ${P}_n(\Omega)$ be expression (\ref{exprbgpn})
but with the $2\pi$ in the
delta function replaced by a new
variable $\Omega$, and let
$\tilde{P}_n(s)$ be its Laplace transform 
with Laplace variable $ns$.
Here we pass to
a ``canonical ensemble'' of angles with weight 
$\exp[-ns\sum_{m=1}^n(\beta_m+\gamma_m)]$. 
We define ${\mathbb H}(s)$ by writing 
$\tilde{P}_n(s) = \int\dd\beta\dd\gamma\,\, \ee^{-{\mathbb H}(s)}$.
The evaluation of this integral constitutes the real problem:
it is on all possible {\it shapes} $(\beta,\gamma)$
of a Voronoi cell with average midpoint distance scaled to unity.
\vspace{5mm}

\noindent {\bf 3. Asymptotic expansion}
\vspace{2mm}

Anticipating that we will be able to identify
a zeroth order problem
we write ${\mathbb H}(s) = {\mathbb H}_{\,0}(s)+{\mathbb V}$.
Hence
\beq
\int\dd\beta\dd\gamma \,\,\ee^{-{\mathbb H}(s)} =
\la\ee^{-{\mathbb V}} \rao
\int\dd\beta\dd\gamma \,\,\ee^{-{\mathbb H}_{\,0}(s)}\,,
\label{spliH0V}
\eeq
where $\la\ldots\rao$ is the average with weight
$\exp[-{\mathbb H}_{\,0}(s)]$.
The appropriate choice of ${\mathbb H}_{\,0}$, 
if there is one at all,
depends delicately on how 
the variables of integration in (\ref{spliH0V})
are assumed to scale with $n$.
We certainly expect that for large $n$ only small
$\beta_m$ and $\gamma_m$ will contribute, but 
the exact scaling of these angles with $n$
is not {\it a priori\,} evident.
The calculation bears out that
\beq
\ee^{-{\mathbb H}_{\,0}(s)} = \pi^{-n}\prod_{m=1}^n 
(\beta_m+\gamma_m)\exp[-ns(\beta_m+\gamma_m)]
\label{defmathH0}
\eeq
is the right choice for ${\mathbb H}_{\,0}$.

We now transform to another set
of angles, {\it viz.}
$\xi_m=\beta_m+\gamma_m$ and $\eta_m=\gamma_m+\beta_{m+1}$
where $m=1,\ldots,n$ and
$\beta_{n+1}\equiv\beta_1$.
Clearly $\xi_m$ is the angle between two
successive midpoint vectors and $\eta_m$ 
the one between two successive vertex vectors. 
Since the set $\{\xi_m,\eta_m\}$ does not fix the relative angle 
between the systems of midpoint and of vertex vectors, 
we complete it by one of the original angles, $\beta_1$. 
Setting $\Phi_m=\sum_{\ell=1}^m\xi_\ell$ and
$\Psi_{m}=\beta_1+\sum_{\ell=1}^{m-1}\eta_\ell$ we obtain 
for $m=1,\ldots,n$ the inverse relation 
\beq
\beta_m = \Psi_{m}- \Phi_{m-1}, \quad
\gamma_m = \Phi_{m} - \Psi_{m}\,.
\label{inversebgxy}
\eeq

The ``zeroth order'' problem is constituted by 
the integral that is left in the RHS of (\ref{spliH0V}) when
one sets ${\mathbb V}=0$. It can be solved when
reformulated in terms of the $\xi_m$ and $\eta_m$.
Since $\xi_m, \eta_m >0$ and
$\sum_m\xi_m=\sum_m\eta_m=2\pi$, they must scale as 
$\xi_m=n^{-1}x_m$ and $\eta_m=n^{-1}y_m$. 
In the limit $n\to\infty$ the $x_m$ and $y_m$
may be integrated from $0$ to $\infty$ and
the problem of the nested
integrations disappears. One finds
$\int\!\dd\beta\dd\gamma\,
\exp[-{\mathbb H}_{\,0}(s)]=(3n-2)!\,\pi^{-n}(ns)^{-3n+1}/(2n)!$,\,\, 
up to contributions that vanish exponentially with $n$. 
Let $P_n^{(0)}(\Omega)$ be the Laplace inverse of this zeroth order 
expression; one readily finds
$P_n^{(0)}(2\pi)=(8\pi^2)^n/[4\pi^2(2n)!]$.\,\,
The inverse Laplace integral 
has a saddle point at $s=s_{\rm c}=3/\Omega$;
if we can show that for $n\to\infty$ the prefactor
$\la\exp(-{\mathbb V})\rao$ in Eq.\,(\ref{spliH0V}) 
has a finite nonzero value in $s=s_{\rm c}$, 
then it follows that
$p_n=\la\exp(-{\mathbb V})\ra_{3/2\pi} P_n^{(0)}(2\pi)$.

\begin{figure}
\begin{center}
\scalebox{.55}
{\includegraphics{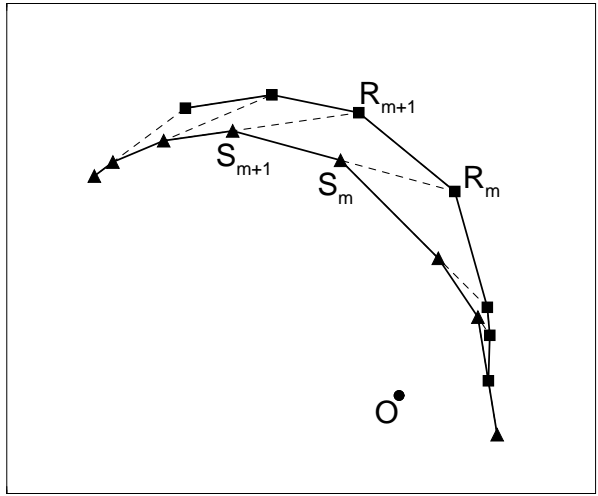}}
\end{center}
\vspace{3mm}
\caption{{\small Schematic picture of the two random walks
associated with the perimeter of a Voronoi cell.
The walk joining the vertices $\vecS_m$ is the actual
perimeter; the other walk joins the midpoints $\vecR_m$.
In the limit $n\to\infty$
the increments of the polar angle along each walk
are independent random variables distributed according to
Eq.\,(\ref{expruv}). The perimeter segments are on the perpendicular
bisectors (dashed lines).}}
\label{fig1}
\end{figure}

We pause for a few comments.
The zeroth order calculation has a useful
corollary: in the limit $n\to\infty$
the scaled angles $x_m$ and $y_m$ become independent
random variables with laws $u(x_m)$ and $v(y_m)$ given by
\beq
u(x)=\pi^{-2}x\,\ee^{-x/\pi}, \qquad
v(y)=(2\pi)^{-1}\ee^{-y/(2\pi)}\,.
\label{expruv}
\eeq
It follows, again for $n\to\infty$, 
that $\{\Phi_m\}_{m=0}^n$ and $\{\Psi_m\}_{m=0}^n$ 
are two mutually independent one-dimensional
random walks, each with a drift of
$2\pi/n$ per step and with independent increments;
they describe how the polar coordinate of the 
midpoint vector 
$\vecR_m=(R_m,\Phi_m)$ 
and of the vertex vector 
$\vecS_m=(S_m,\Psi_m)$, 
respectively, increases as one progresses along the perimeter.
These angular random walks {\it drive}
the radial coordinates $R_m$ and $S_m$ and couple them together,
which leads to the double-stranded {\it two}-dimensional 
walk depicted in Fig.\,\ref{fig1}. 
\vspace{2mm}

Let us now pursue the analysis. The validity of the 
zeroth order results hinges on our ability to
prove that the first factor in Eq.\,(\ref{spliH0V}), 
$\la\exp(-{\mathbb V})\rao$, exists and is subdominant in the 
large $n$ expansion of $\log p_n$.
Actually, everything has been so prearranged that
this factor is of order $n^0$. 
Moreover, it has an interesting structure.

The regular $n$-sided polygon
is an obvious point of symmetry in phase space.
DI obtained an exact lower bound for $p_n$ by expanding
the vertex vectors $\vecS_m$ about their regular polygon positions.
This procedure preserves, however, the long range polygonal order.
Here we will advance as in the theory of
elasticity, where one considers the deviations 
not of the positions, but of the interatomic
distances, from their ground state values.
Hence we expand ${\mathbb V}$ in powers of 
$\delta x_m=x_m-\bar{x}$ and $\delta y_m=y_m-\bar{y}$ 
(where $\bar{x}=\bar{y}=2\pi$).
This procedure allows, in principle, 
for large deviations from the regular $n$-sided polygon.

Terms from various sources
contribute to ${\mathbb V}$ and have to be examined one by one.
The scaling of the
$\xi_m$ and $\eta_m$ and their asymptotic independence
determine how the other variables scale;
Eq.\,(\ref{inversebgxy}) implies, in particular, 
that the original angles scale as
$\beta_m=n^{-\half}b_m$ and $\gamma_m=n^{-\half}c_m$.
One finds that ${\mathbb V}={\mathbb V}_0
+n^{-\frac{1}{2}}{\mathbb V}_{1}+\ldots$\,. 
Hence $\la\exp(-{\mathbb V})\rao = \la\exp(-{\mathbb V}_0)\rao\,
[1+{\cal O}(n^{-\frac{1}{2}})]$.
It appears that ${\mathbb V}_0$ is quadratic in the $\delta x_m$ 
and $\delta y_m$ with contributions coming from (i)
the ${\cal O}(n^{-1})$ terms in the expansion  
${A}=1+n^{-1}{A}_1+\ldots$, and (ii) the product
$\prod_{m=1}^n \rho_m^2 T_m/(\beta_m+\gamma_m)$.
In terms of the
Fourier components $\hat{z}_q=
(2\pi n^{\frac{1}{2}})^{-1}\summ \ee^{2\pi{\rm i} qm/n} \delta z_m$, where
$z=x, y$ and $q$ is integer, one finally finds, with $s=3/(2\pi)$,
\beq
{\mathbb V}_0 =
\sum_{q \neq 0}\,
\big[ (q^{-2} + 2q^{-4})
(\hat{x}_{q}-\hat{y}_{q})
(\hat{x}_{-q}-\hat{y}_{-q}) 
\,+\, 2q^{-2}(\hat{x}_{q}-\hat{y}_{q})\hat{y}_{-q} \big]\,,
\label{mathbbV0FT}
\eeq
where $q$ runs through all nonzero integers.
The decay of the Fourier coefficients as $\sim q^{-2}$ for large $q$
is characteristic of interacting Coulomb charges.
Since the $\hat{x}_q$ and $\hat{y}_q$
are sums of independent variables, they are Gaussian 
distributed with variances determined by (\ref{expruv}).
We can show that the quantity $\exp(-{\mathbb{V}}_0)$ 
may be averaged by Gaussian integration:
in terms of Coulomb charges this amounts to a
Debye-\-H\"uckel approximation, which becomes exact in the high
temperature limit; here the inverse
``temperature'' is $1/n$ and the exactness
of our procedure follows \cite{HJH05}
from a Hubbard-Stratonovich transformation.
Gaussian integration then leads to
$\la\exp(-{\mathbb{V}}_0)\ra_{3/2\pi}=C$ where 
$C = \prod_{q=1}^\infty\,( 1-q^{-2}+4q^{-4} )^{-1}\,=\,0.344...$
Hence the finite size correction factor $C$
is the partition function of the 
deformational modes of the perimeter with wavelength $2\pi/q$, 
the main contributions coming from the long
wavelengths.
\vspace{5mm}

\noindent {\bf 4. Results}
\vspace{2mm}

Upon combining the zeroth and first order result we
conclude that the probability ${p}_n$
for a Voronoi cell to be $n$-sided behaves as
\begin{equation}
{p}_n = \frac{C}{4\pi^2}\,\frac{(8\pi^2)^n}{(2n)!} 
\,\big[1+{\cal O}(n^{-\frac{1}{2}})\big],\qquad n\to\infty,
\label{resultpn}
\end{equation}
which may also be written as the asymptotic equality
\beq
\log{p}_n \simeq -2n\log n + n\log (2\pi^2\ee^2) 
- \tfrac{1}{2}\log ( 2^6\pi^5 C^{-2}n ). 
\label{resultlogpn}
\eeq
We compare this to existing work.
DI assume an expansion similar to (\ref{resultlogpn}) and
provide strong analytic an numerical evidence
for the coefficient of the $n\log n$ term 
to be equal to $-2$. This has now been confirmed.
Furthermore, DI prove that the term proportional to $n$ 
has a coefficient larger than $\log(\pi^2/\ee)=1.289...$,
and from their numerical data estimate 
that it is actually between 4 and 5.
Our value $\log(2\pi^2\ee^2)=4.982...$ 
satisfies DI's exact inequality, as of course it had to,
and is just inside the range of values that these authors
judged likely.

\begin{figure}
\begin{center}
\scalebox{.75}
{\includegraphics{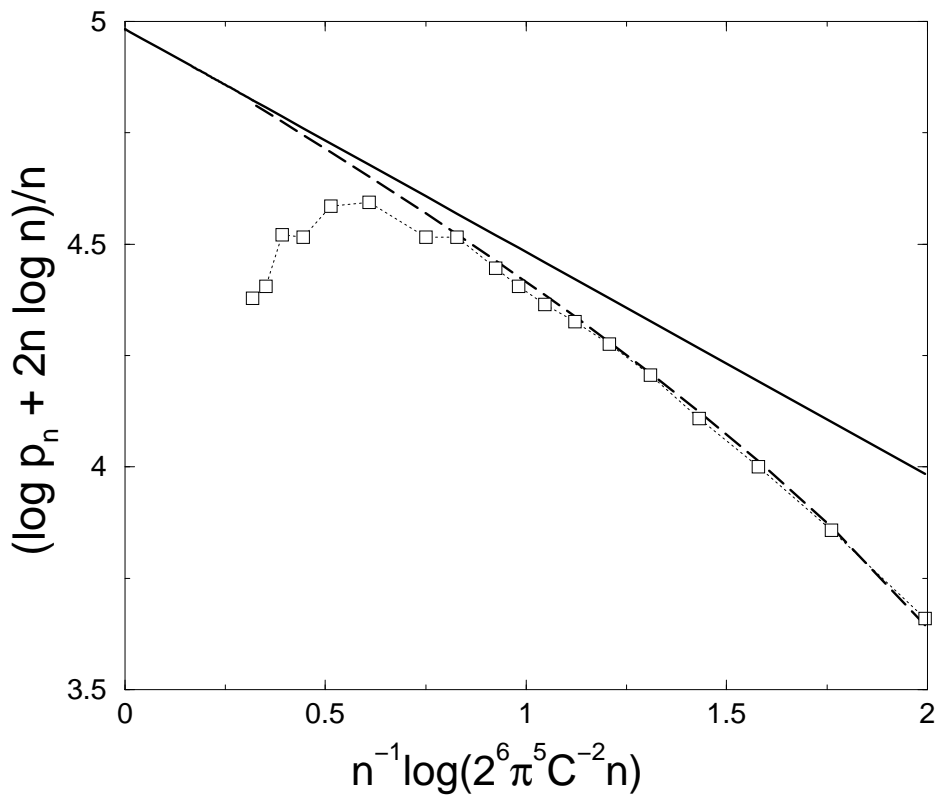}}
\end{center}
\caption{{\small Straight solid line: 
the asymptotic formula, Eq.\,(\ref{resultlogpn}).
Open squares: Monte Carlo data by Drouffe and Itzykson (Ref.\,\cite{DI84},
Table 1).}
Dashed line: Eq.\,(\ref{resultlogpn}) with an additional term 
$an^{-\frac{1}{2}}$, where we adjusted $a=-0.24$ to fit the data.}
\label{fig2}
\end{figure}
 
In Fig.\,\ref{fig2} our asymptotic result Eq.\,(\ref{resultlogpn})
is shown as a straight solid line of slope $-\frac{1}{2}$. It is
compared to the Monte Carlo data of Ref.\,\cite{DI84} (open squares)
in the range $50\geq n\geq 7$
(the abscissa value of $2$ corresponds
very nearly to $n=7$).
The first term not displayed in the asymptotic series
of Eq.\,(\ref{resultlogpn}) will be of the form $an^{-\frac{1}{2}}$;  
if, venturing at this point beyond our analytic results, we
choose its amplitude to fit the data, very good agreement
(the dashed line) with the Monte Carlo results
is obtained for $a=-0.24\pm 0.02$. 
We believe that the downward trend in the DI data for $n\gsim 30$
(abscissa $\lsim 0.75$)
is due to the great difficulty of such simulations;
in that region the authors estimated their error bars 
to be at least of the order of the data themselves. 

Consistency of the asymptotic scaling with 
Eq.\,(\ref{condcosbcosg}) requires that
$\rho_m=1+n^{-\frac{1}{2}}r_m$,
where $r_m$ remains of order $n^0$ as $n\to\infty$.
Hence the midpoint distances $R_m=R\rho_m$ deviate by
order $n^0$ from their average $R$, which itself is typically within 
$n^0$ from $R_{\rm c}$. Therefore the area
of the $n$-sided 
Voronoi cell is sharply peaked around an average equal to
$\pi R_{\rm c}^2=n/(4\nu)$, where we restored the particle density. 
This demonstrates that to leading order Lewis' law holds. 
Its coefficient  $c=\frac{1}{4}$
agrees with DI's observation that $c\approx\frac{1}{4}$.

It is amazing that the analysis of
a problem defined on only a random set of points in the
plane evokes associations with so many phenomena in statistical physics.
A full account \cite{HJH05} of this work
is in preparation. In an
extension of it we will revisit Aboav's law.
\vspace{3mm}

{\it Acknowledgments.}\,\,
The author thanks Pierre Calka for a discussion and for pointing out
some of the mathematical literature on this subject, and
Bernard Jancovici for conversations on Coulomb charges. 
He thanks Jean-Michel Drouffe for correspondence.
A casual remark by Emmanuel Trizac sparked this work off.
The Laboratoire de Physique Th\'eorique of the Universit\'e de Paris-Sud
is associated with the Centre National de la Recherche Scientifique as
research unit UMR 8627.

\end{document}